\documentclass{PoS}
\usepackage{amsmath}
\usepackage{amssymb}
\usepackage{bbold}

\newcommand{\Cnpp}{C_{\frac{1}{2} \rightarrow \frac{1}{2}}^{N}}

\newcommand{\Cvxx}{C_{x \rightarrow x}^{V}}
\newcommand{\Cvyx}{C_{x \rightarrow y}^{V}}

\newcommand{\Cvxy}{C_{y \rightarrow x}^{V}}
\newcommand{\Cvyy}{C_{y \rightarrow y}^{V}}

\title{$J/\psi$	-nucleon scattering in $P_{c}^{+}$ pentaquarks channel}

\ShortTitle{$NJ/\psi$ scattering }

\author{\speaker{Ursa Skerbis} \\
        Jozef Stefan Institute, 1000 Ljubljana, Slovenia\\
        E-mail: \email{ursa.skerbis@ijs.si}}

\author{Sasa Prelovsek\\
        Department of Physics, University of Ljubljana, 1000 Ljubljana, Slovenia \\
        Jozef Stefan Institute, 1000 Ljubljana, Slovenia \\
        Instit\"ut f\"ur Theoretische Physik, Universit\"at Regensburg, D-93040 Regensburg, Germany \\
        E-mail: \email{sasa.prelovsek@ijs.si}}

\abstract{Two pentaquarks $P_{c}^{+}$  were discovered by LHCb collaboration as peaks in the proton-$J/\psi$ invariant mass. We perform the lattice QCD study of the scattering between $J/\psi$ meson and nucleon in the channels with $J^{P}=\frac{3}{2}^{+},\frac{3}{2}^{-}, \frac{5}{2}^{+}, \frac{5}{2}^{-}$, where $P_{c}^{+}$ was discovered. This is the first lattice simulation that reaches the energies $4.3-4.5~$GeV where pentaquarks reside. The higher partial waves $L>0$ are also explored for the first time. In this study we consider the single-channel approximation for scattering of $NJ/\psi$. Energies and eigenstates are extracted for the $NJ/\psi$ system at the zero total momentum for all six irreducible representations of the lattice irreducible representation. No significant energy shifts are observed. The number of eigenstates agrees with the number expected from non-interacting limit for scattering.  This could possibly indicate that the $P_{c}$ resonances seen in experiment are a consequence of a coupling of the $NJ/\psi$ channel with other two-hadron channels. }

\FullConference{The 36th Annual International Symposium on Lattice Field Theory - LATTICE2018\\
		22-28 July, 2018\\
		Michigan State University, East Lansing, Michigan, USA.}

\begin{document}

\section{Introduction}

Two peaks in a proton-$J/\psi$ invariant mass were discovered in $2015$ by LHCb \cite{LHCb_Pc_2015}. This discovery was later confirmed by a model independent study in $2016$ by the same collaboration \cite{2016}. 
Two resonances were observed, the broader with width $\Gamma=205  \text{ MeV}$ and mass $M \approx 4380 \text{ MeV}$  and the narrower with $\Gamma=40 \text{ MeV}$ and mass $M \approx 4450 \text{ MeV}$. These resonances were later identified as hidden charm pentaquarks $P_{c}$ with minimal flavor structure $uudc\bar{c}$.  LHCb found  the best fit for spin-parity assignments $(J_1^{P_1},J_2^{P_2})=(\frac{3}{2}^- ,\; \frac{5}{2}^+)$, while acceptable solutions are also found for additional cases with the opposite parity, either ($\frac{3}{2}^+ , \;\frac{5}{2}^-$) or
($\frac{5}{2}^+,\; \frac{3}{2}^-$).  $P_c$ resonances can strongly decay to a nucleon and a charmonium as well as to a charmed baryon and charmed meson.

At present there is no knowledge on $P_c$ resonances based on the first-principle lattice QCD. The lattice simulations of systems with flavor $\bar{c}cuud$ have never reached energies where the pentaquarks reside. Previous dynamical \cite{Jpsi_nucleon_HALQCD_method} and quenched   \cite{Jpsi_nucleon_HALQCD_method} studies  \cite{Kawanai:2010ev} presented  results for  $NJ/\psi $ and $N\eta_c$  potentials and phase shifts  in s-wave using HALQCD method in one-channel approximation. These were extracted up to the energies $0.2~$GeV above threshold. An attractive interaction was found in all channels explored, but not attractive enough to form bound states or resonances.  The hadroquarkonium picture was considered in \cite{1608.06537_Alberti2016}, where the static $\bar cc$ potential  $V(r)$ was extracted for $m_c\to \infty$ as function of distance $r$  in the presence of the nucleon. The potential is found shifted down only by a few MeV due to the presence of the nucleon. 

This is the  first lattice simulation of $NJ/\psi$ scattering  that reaches the energies $4.3-4.5~$GeV where pentaquarks reside. We explore partial wave   $L=0$ and for the first time also $L>0$. The aim is to explore the fate of pentauqrak in one-channel approximation, where $N J/\psi$ is decoupled from other two-hadron channels.  We therefore perform a simulation of  $NJ/\psi$ scattering in one-channel approximation in order to find whether $P_c$ features in the spectrum in this case. The detailed presentation of the study is given in   \cite{Skerbis2018}, together with analogous simulation of the $N\eta_c$ channel. 

\begin{figure}
	\centering
	\includegraphics[width=0.55\linewidth]{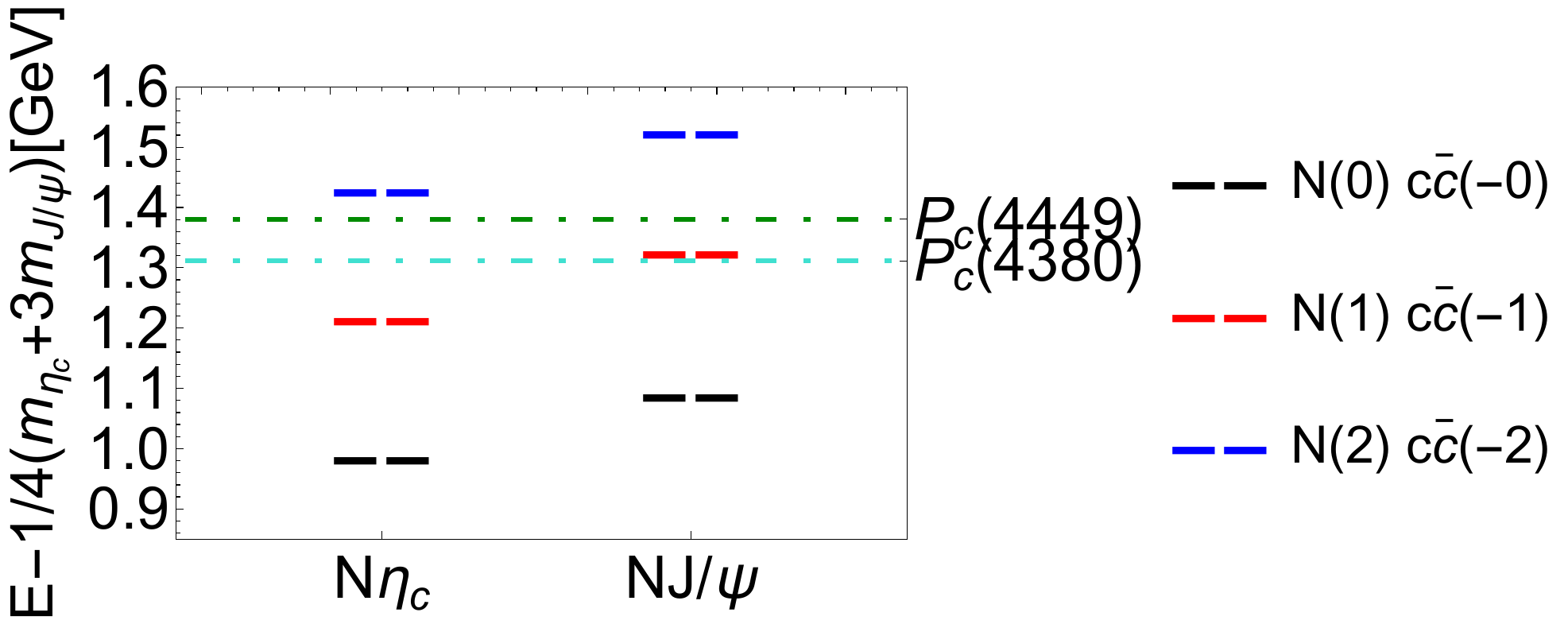}
	\caption{Non-interacting energies for the nucleon-charmonium system on our lattice  (\ref{eq:non_int_energy}). Green and turquoise dash-dotted lines are added at $P_{c}$ masses. }
	\label{fig:energijelattice}
\end{figure}

The energy spectrum of the $NJ/\psi$ system in the non-interacting limit is an important reference case for scattering studies. The momenta $\bold{p}=\bold{n}\frac{2 \pi}{{\tt L}}$ of each hadron are discrete due to periodic boundary conditions of fermions in space on the lattice. The non-interacting energies of the nucleon-meson system are
\begin{equation}
\label{eq:non_int_energy}
E_{n.i.}=E_{N}(\bold{p})+E_{V}(-\bold{p}),~~\bold{p}=\bold{n}\frac{2 \pi}{{\tt L}},~~ \Delta E= E -E_{n.i.}
\end{equation}
with $\bold{n} \in \mathbb{N}^{3}$. The  $E_{H=N,V}(p)$  \footnote{Due to simplicity, $p$ will be used instead of $\bold{p}$ from here on.}  are single hadron energies measured for different momenta on our lattice, they satisfy $E_{H=N,V}(p)=\sqrt{m_{H=N,V}^2+ p^2}$ in the continuum.  Non-interacting energies $E_{n.i.}$ are shown in Figure \ref{fig:energijelattice}, together with  experimental  masses of  $P_{c}$ resonances.
In order to capture the region of both resonance, $NJ/\psi$ channel is explored up to $p^2 \leq 2$. The aim is to determine eigen-energies of the $NJ/\psi$ system. Energies $E_n$ are compared to non-interacting energies $E_{n.i}$ in search for  the energy shift $\Delta E$. Significant non-zero energy shift $\Delta E \neq 0 $ or an additional eigenstate could indicate the presence of a resonance state in the  system \cite{Luscher1991531}. 

\section{Single hadron operators}
\label{sec:single_had}
In order to determine non-interacting energies of the $NJ/\psi$ system, the energies of nucleon and meson are computed separately.  We used $3$ standard nucleon interpolators (Eq. \ref{eqn:op_nucleon}) and $2$ standard mesonic interpolators (Eq. \ref{eqn:op_meson})  for each  value of relative momenta $p$. 
\begin{eqnarray}  
\label{eqn:op_nucleon}
& N(\vec{p},t)=\sum_{\vec{x}}\epsilon_{abc} P^{+} \Gamma^{1} u(\vec{x},t) (u^{T}(\vec{x},t)\Gamma^{2}d(\vec{x},t)) e^{i\vec{p}\vec{x}}, ~ (\Gamma^{1},\Gamma^{2}):     (\mathbb{1},C\gamma_{5}) ,\ (\gamma_{5},C),\ (\mathbb{1},\imath \gamma_{4}C\gamma_{5})   \\
\label{eqn:op_meson}
& \quad V(\vec{p},t)=\sum_{\vec{x}} c(\vec{x},t)\Gamma \bar{c}(\vec{x},t) e^{i\vec{p}\vec{x}}  ~~~~~~~~~ \Gamma:\   \gamma_{i} ,  \gamma_{i} \gamma_{5}, \ i=x,y,z   
\end{eqnarray}

\section{Two-hadron operators, expected degeneracy and construction of two hadron correlators }
\label{sec:two_hadron_op_to_corr}
The operators for scattering of particles with spin were already employed within our previous work \cite{operators_2017}, where all explicit expressions  for operators of form $O \approx N(p)V(-p)$ for $p^2 \leq 1$ are given.  We employ operators in Partial wave method \cite{Callat_partial_wave, operators_2017}
\begin{equation}
\label{eqn:partial_wave}
O^{|p|,J,m_J,L,S}=\sum_{m_L,m_S,m_{s1},m_{s2}} C^{Jm_J}_{Lm_L,Sm_S}C^{Sm_S}_{s_1m_{s1},s_2m_{s2}}  
\sum_{R\in O} Y^*_{Lm_L}(\widehat{Rp}) N_{m_{s1}}(Rp)M_{m_{s2}}(-Rp)~.~~~~
\end{equation}
These are subduced to the chosen  irrep $\Gamma$ (Eq. \ref{eqn:subduction})  using subduction coefficients  ${\cal S}^{J,m_J}_{\Gamma,r}$ from \cite{PolSpinEdwardsDudek2011}.
\begin{equation}
\label{eqn:subduction}
O_{|p|,\Gamma,r}^{[J,L,S]}=\sum_{m_J}  {\cal S}^{J,m_J}_{\Gamma,r} O^{|p|,J,m_J,L,S}.
\end{equation}

On the lattice, the operators $O^{J,m_J}$ form a reducible representation with respect to the lattice group $O_h$. One has to employ operators which transform according to irreducible representations $\Gamma^P$. Those are listed in Table \ref{tab:subduction_depending_on_J}, where several $J^P$ contribute to a given irrep $\Gamma^P$. 

\begin{table}[h!]
	\centering
	\begin{tabular}{c|c}
		irrep $\Gamma^P$ & $J^P$ \\
		\hline
		$G_{1}^\pm$ & $\frac{1}{2}^\pm$ ,  $\frac{7}{2}^\pm$ \\
		$G_{2}^\pm$ & $\frac{5}{2}^\pm$ ,  $\frac{7}{2}^\pm$ \\	
		$ H^\pm$ & $\frac{3}{2}^\pm$ , $\frac{5}{2}^\pm$ ,  $\frac{7}{2}^\pm$ \\	
	\end{tabular}
	\caption{Irreducible representations $\Gamma^P$ of the discrete lattice group $O_h$, together with a list of $J^P$ that a certain irrep contains.   }
	\label{tab:subduction_depending_on_J}
\end{table}

$P_{c}$ states with $J=3/2^\pm $ or $5/2^\pm$ could  be seen in irreps $G_{2}^{\pm} \text{ or } H^{\pm}$.
A simple example of the  $NJ/\psi$ operator at $p=0$ that transform according to the $H^-$  irrep is
\begin{equation}
\label{eq:anihilation_O_Hm}
O^{H^-,r=1}_{(J,L,S)=(\frac{3}{2},0,\frac{3}{2})}(0)=N_{\frac{1}{2}}(0)\left(V_x(0)-i V_y(0)\right) .
\end{equation}

 In a non-trivial case, relations \ref{eqn:partial_wave} and \ref{eqn:subduction} lead to multiple linear-dependent operators. For each irrep one can find linearly independent basis of operator-types. The employed  operator-types  can be found in appendix of \cite{Skerbis2018}.  All remaining operators can be written as a linear combination of these. 

In the non-interacting limit, one expects several degenerate $N(p)V(-p)$ eigenstates  for most of $J^P$ (or irreps) and relative momenta $p\!>0\!$. 
In the continuum, different combinations of $(L,S)$   lead to a given $J^P$ ($|L-S|\leq J\leq |L+S|$)  due to the non-zero spins of the scattering particles.  The linearly independent combinations $(L,S)$ represent linearly independent eigenstates, so each of them should feature as an independent eigenstate in the spectrum.   On the lattice,  also different spins $J^P$ can contribute to a given irrep $\Gamma^P$ as listed in Table \ref{tab:subduction_depending_on_J}.  Linearly independent combinations coresponding to $(J^P,L,S)$, that subduce to given irrep $\Gamma^P$, now present linearly-independent eigenstates. The numbers of these states are  summarized in Table \ref{tab:num_of_operators} -  those are the number of degenerate eigenstates  in a given row of irrep in the non-interacting limit.  The number of linearly independent operator-types is also equal to the number of degenerate eigenstates.
\begin{table}
	\centering
	\begin{tabular}{c|ccc||}
		$\text{irrep}$ &  \multicolumn{3}{c}{$N(p)J/\psi(-p)$}   \\ 
		
		& $p^2=0$  & $p^2=1$  & $p^2=2$  \\ 
		\hline 
		\hline 
		$G_{1}^{+}$  & $0$ & $2$ & $3$ \\ 
		\hline 
		$G_{1}^{-}$ & $1$ & $2$ & $3$ \\ 
		\hline 
		$G_{2}^{+}$ & $0$ & $1$ & $3$ \\ 
	\end{tabular}
	\begin{tabular}{c|ccc}
		$\text{irrep}$ &  \multicolumn{3}{c}{$N(p)J/\psi(-p)$}  \\ 
		& $p^2=0$  & $p^2=1$  & $p^2=2$  \\ 
		\hline 
		\hline 
		$G_{2}^{-}$ & $0$ & $1$ & $3$ \\ 
		\hline 
		$H^{+}$  & $0$ & $3$ & $6$ \\ 
		\hline 
		$H^{-}$  & $1$ & $3$ & $6$ 
		
	\end{tabular}
	\caption{ The number of the expected degenerate eigenstates for each row of irrep (in non-interacting limit). Each of those linearly independent eigenstates should appear in the spectrum.  This number is equal to the  number of the linearly-independent operator-types. }
	\label{tab:num_of_operators}
\end{table} 

\begin{figure}[h!]
	\centering
	\includegraphics[height=0.13\textheight]{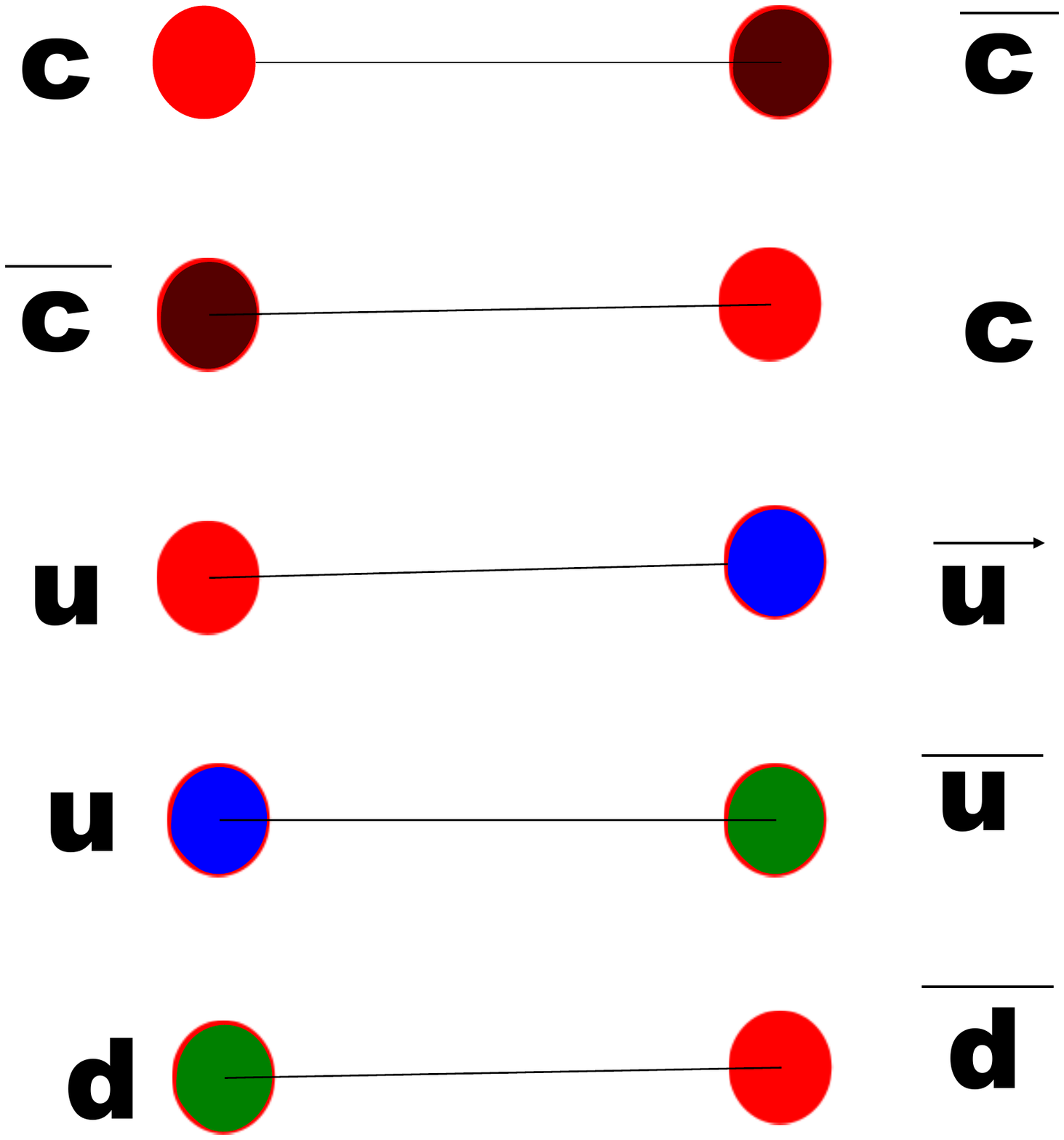}
	\includegraphics[height=0.13\textheight]{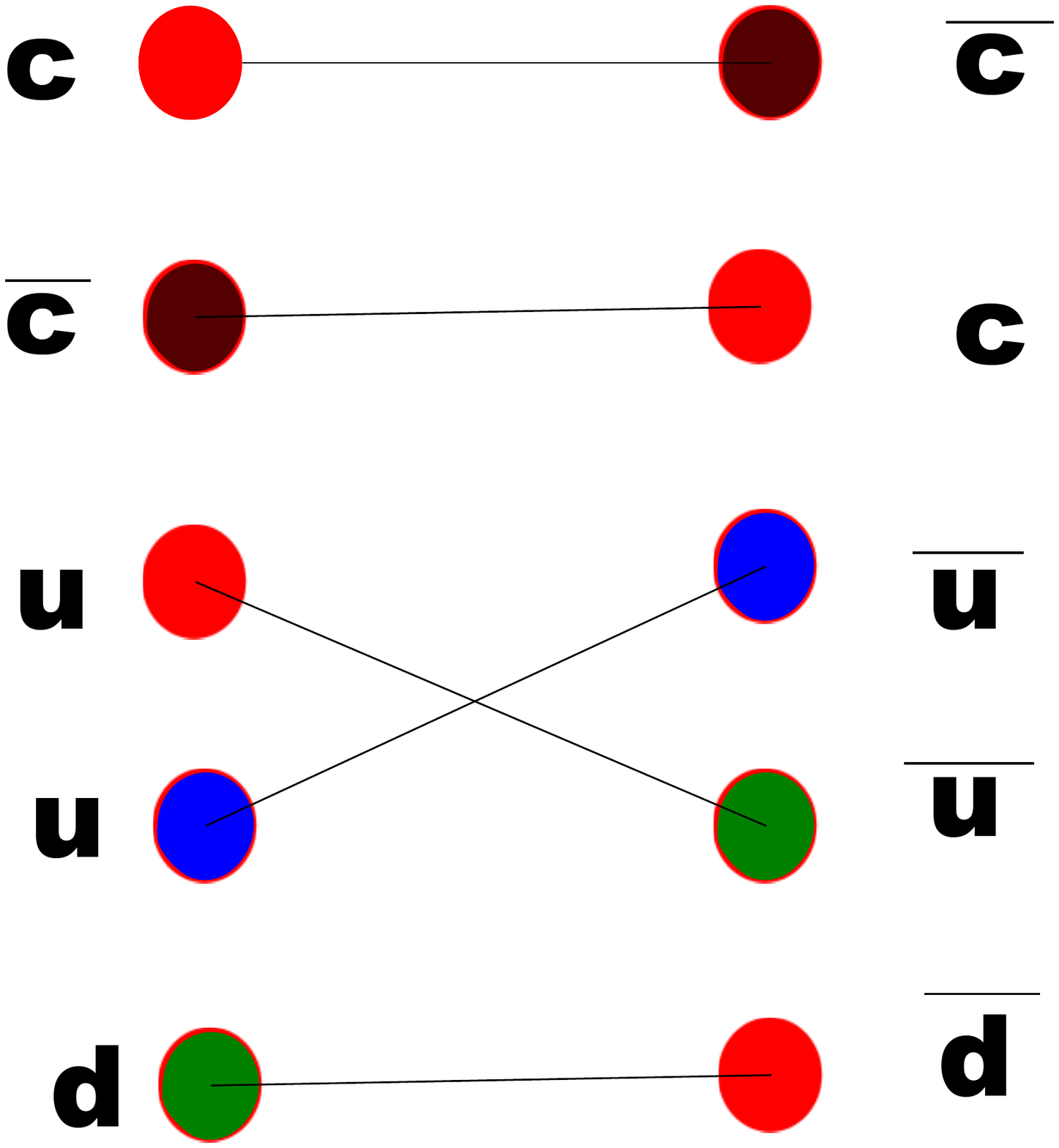}
	\caption{Wick contractions considered in our simulation for one-channel approximation.}
	\label{fig:pccontr}
\end{figure}

In the elastic approximation there is no contraction connecting $J/\psi$ and $N$ interpolators, as shown in Figure \ref{fig:pccontr}.
Therefore a single-hadron correlation function can be simulated separately and later combined to the two-hadron correlation functions. 
An example of two-hadron correlator corresponding to operator \ref{eq:anihilation_O_Hm} at the sink and its creation operator $\bar{O}^{H^-,r=1}_{(J,L,S)=(\frac{3}{2},0,\frac{3}{2})}$  at the source is given in Equation \ref{eqn:corr_function_Hm} \begin{eqnarray}
\label{eqn:corr_function_Hm}
&C^{VN;H^-}_{(J,L,S)=(\frac{3}{2},0,\frac{3}{2})}(0)= \Cnpp\Cvxx -i \Cnpp \Cvyx+ i \Cnpp \Cvxy + \Cnpp \Cvyy,  \\
& \quad C^{H}_{pol_{src}\rightarrow pol_{snk}} =\langle\Omega|H_{pol_{snk}}\bar{H}_{pol_{src}}|\Omega\rangle, ~~  H=N,V. \nonumber
\end{eqnarray}
Similarly, all   two-hadron correlators in our study can be expressed in terms of 
$ \langle0|N_{m_s'}(p')\bar{N}_{m_s}(p)|0\rangle$, $ \langle0|V_{i'}(p')V_{i}^\dagger (p)|0\rangle$ and $ \langle0|P(p') P^\dagger(p)|0\rangle$, which were pre-computed for all combinations of $p',p=0,1,2$, $i,i'=x,y,z$ and $m_s,m_s^\prime=1/2,-1/2$. 
 
\section{Lattice setup}
All simulations were performed on $N_f=2$ ensemble with parameters listed in table \ref{tab:latt},  that was generated in context of the work \cite{Latt_data_Hasenfratz_2008,Latt_data_Hasenfratz_2_2008}.  
\begin{table}[h!]
	\centering	
	\begin{tabular}{c c c c c c }
		$N^{3} \times N_{T}$ & $\beta$ & $a\text{[fm]}$ & ${\tt L}\text{[fm]}$  & $\# \text{config}$ & $m_{\pi}\text{[MeV]}$ \\
		\hline
		$16^{3} \times 32$ & $7.1$ & $0.1239(13)$ & $1.98(2)$  & $281$ & $266(3)$ 
	\end{tabular}
	\caption{ Parameters of the lattice ensemble. }
	\label{tab:latt}
\end{table}

Wilson-Clover action is used for light quarks while for charm quarks  Fermi lab approach is employed.  Full distillation was used for quark smearing. $48$ eigenvectors were used for smearing of light quarks in  nucleon, while charm quarks in charmonium were smeared  with use of $96$ eigenvectors. 
 
\section{Results} 

 Resulting eigen-energies for single and two hadron system are presented. All results are obtained from the correlated one exponential fits and the errors are calculated using jack-knife method. 
\subsection{Individual energies of $N$ and $J/\psi$}
The energies of  nucleon and $J/\psi$ meson for  various momenta $p^2=0,1,2$ are given in Table \ref{tab:single_hadron_fit}.  
Those are needed to determine (\ref{eq:non_int_energy}).
\begin{table}[h!]
	\centering
	\begin{tabular}{c c | c c   c ||}
		\text{particle} &$p^{2}$ & $E_na$ &  $\sigma_{E_n}a$& \text{fit range} \\
		\hline
		\hline
		$N$ &  $0$& $0.701$ & $0.019$ &  $[6,9]$ \\ 
		&  $1$& $0.769$ & $0.028$& $[7,10]$ \\ 
		& $2$& $0.849$ & $0.054$ & $[7,9]$ \\ 
	\end{tabular}
	\begin{tabular}{c c | c c   c}
		\text{particle} &$p^{2}$ & $E_na$ &  $\sigma_{E_n}a$& \text{fit range} \\
		\hline
		\hline
		$J/\psi$& 	$0$& $1.539$ & $0.001$ & $[10,14]$ \\ 
		& 	$1$& $1.576$ & $0.001$ &$[10,14]$ \\ 
		& 	 	$2$& $1.613$ & $0.001$ & $[9,12]$ \\ 
	\end{tabular}
	\caption{ Fitted energies for single hadrons. }
	\label{tab:single_hadron_fit}
\end{table}

\subsection{ $NJ/\psi$ channel}
Eigen-energies of $NJ/\psi$ system were extracted from the correlation matrices using GEVP.  This big correlation matrices give rather noisy eigenvalues, therefore we restricted our analysis to a smaller subset, where each operator-type is represented by two operators:   both meson operators (\ref{eqn:op_meson}) and the first nucleon operator (\ref{eqn:op_nucleon}).  Energies are obtained from  the eigenvalues using  the correlated one-exponential fits, while their errors are calculated using jack-knife method.  All fits were performed for $t=[7,10]$. 

\begin{figure}
	\centering
	\includegraphics[width=0.8\linewidth]{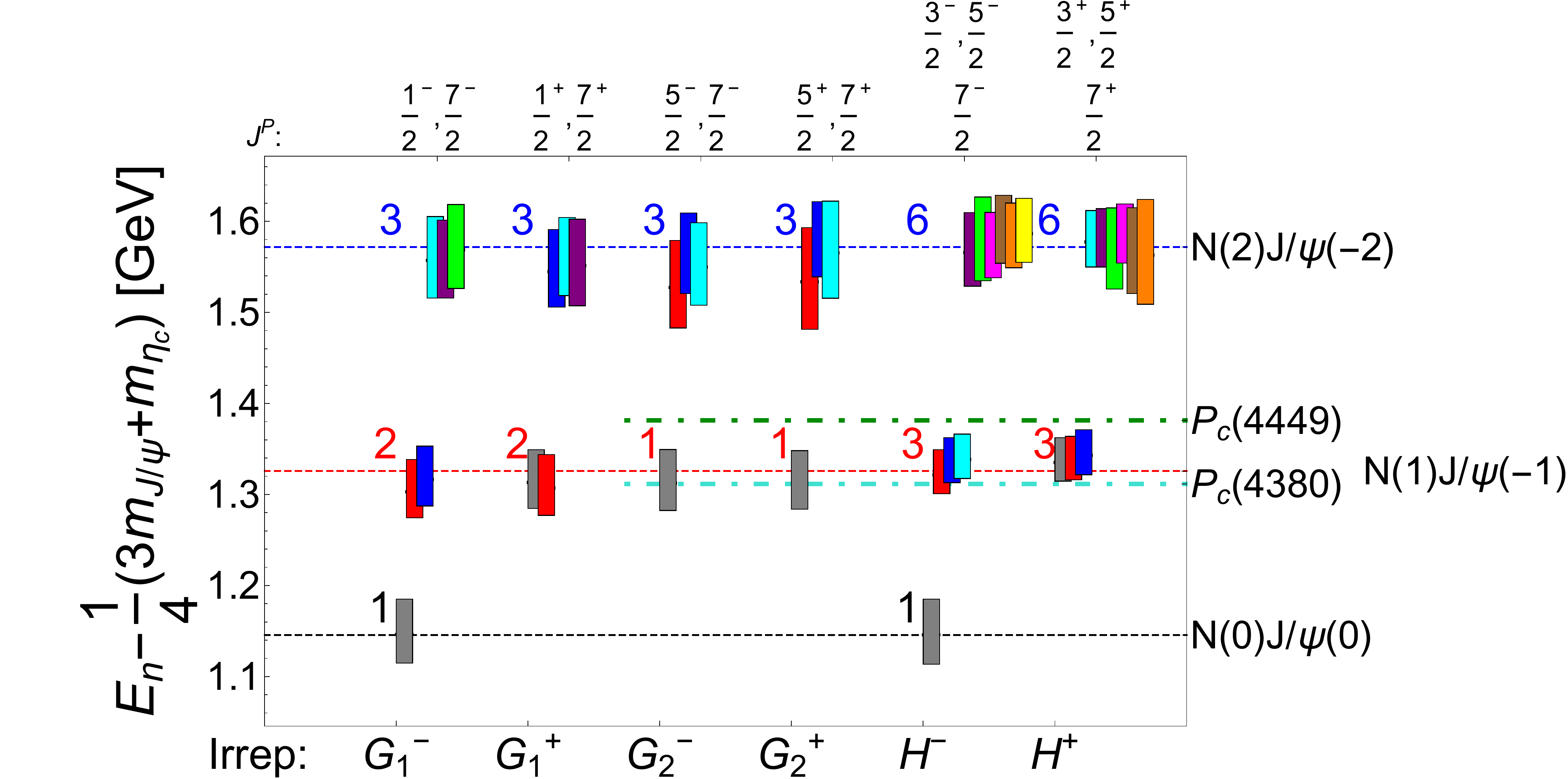}
	\caption{Energy spectrum  of  $NJ/\psi$ system in all $6$ irreps of $O_{h}^{2}$. Dashed lines are non-interacting $NJ/\psi$ energies (Eq. \ref{eq:non_int_energy}).  Green and turquoise dash-dotted lines are experimental values of $M_{P_c}$. Center of rectangle is eigen-energy $E_{n}$ and its height corresponds to $2\sigma_{E}$.  The number of expected states in non-interacting case is added to a figure on a upper left side at each data set.}
	\label{fig:plotmainoriginalcombination}
	\includegraphics[width=0.50\linewidth]{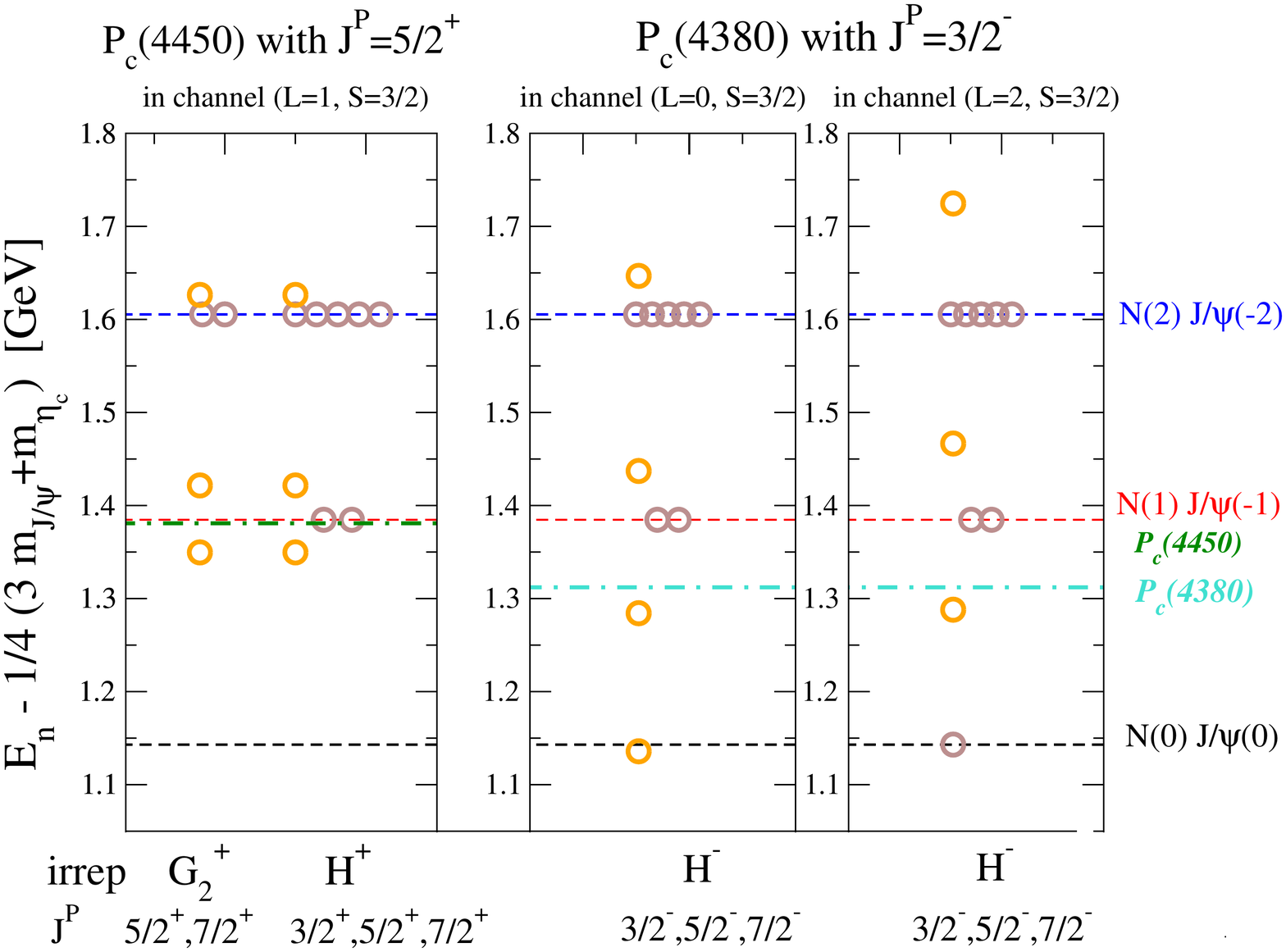}
	\caption{The energies of eigenstates in a scenario with a Breit-Wigner-type $P_c(4450)$ or $P_c(4380)$ resonances, assuming that it is coupled only to $NJ/\psi$ channel and decoupled from other two-hadron channels. This scenario renders an additional eigenstates near  $M_{P_c}\pm \Gamma_{P_c}$ with respect to the  non-interacting case.}
\label{fig:Pc_analytic}
\end{figure}

The observed spectrum is shown in Fig. \ref{fig:plotmainoriginalcombination}. The energies are compatible with non-interacting ones  (\ref{eq:non_int_energy}) within our errors. We establish all almost-degenerate states expected in the non-interacting limit (Table \ref{tab:num_of_operators}). So our lattice results show no significant energy shift or any additional eigenstate. $P_{c}$ candidate  channels ( $J^{P}=\tfrac{5}{2}^{+} \text{ and } \tfrac{3}{2}^{-}$) are compared to the analytic prediction of eigen-energies  in a scenario with Breit-Wigner-type $P_c(4450)$ or $P_c(4380)$ resonances, shown in Figure \ref{fig:Pc_analytic}.  In a scenario featuring $P_c$  (Figure \ref{fig:Pc_analytic}) we would expect an additional eigenstate (with respect to the non-interacting case) at an energy close  to $M_{P_c}$. These additional eigenstates are not found in our study. We conclude that the scenario based on a Breit-Wigner-type $P_c$ resonances, coupled solely to $NJ/\psi$,  is  not supported by our lattice data. 

\section{ Conclusion}

 We perform a $N_f=2$  lattice QCD simulation of $NJ/\psi$ scattering in the one-channel approximation, where $N$ denotes a proton or a neutron.   The resulting energies of eigenstates  in Figure  \ref{fig:plotmainoriginalcombination}  are compared to the analytic predictions of a scenario with non-interacting $NJ/\psi$ system and a scenario featuring a $P_c$ resonance coupled to a single channel.  We find that the extracted lattice spectra is consistent with the prediction of an almost non-interacting $NJ/\psi$   system within errors of our calculation. The scenario based on a Breit-Wigner-type $P_c$ resonance, coupled solely to $NJ/\psi$,  is  not supported by our lattice data. This might suggest that the strong coupling between the $NJ/\psi$ with other two-hadron channels might be responsible for the existence of the $P_c$ resonances in the experiment.  Future lattice simulations of coupled-channel scattering  are needed to investigate this hypothesis. 

More details on this study can be found in \cite{Skerbis2018}, where  also $N\eta_{c}$ scattering in one-channel approximation is considered.   \\

\vspace{0.1cm}
\textbf{Acknowledgments} 

\vspace{0.1cm}

We thank  A. Hasenfratz, C.B. Lang, L. Leskovec, D. Mohler and  M. Padmanath.      This work was supported by Research Agency ARRS (research core funding No. P1-0035 and No. J1-8137) and DFG grant No. SFB/TRR 55.

\bibliographystyle{JHEP}

\end{document}